# Re-Examination Of Possible Bimodality Of Gallex Solar Neutrino Data

P.A. Sturrock

Center for Space Science and Astrophysics, Varian 302,

Stanford University, Stanford, CA 94305-4060, U.S.A.

e-mail: sturrock@stanford.edu

**Abstract.** The histogram formed from published capture-rate measurements for the GALLEX solar neutrino experiment is bimodal, showing two distinct peaks. On the other hand, the histogram formed from published measurements derived from the similar GNO experiment is unimodal, showing only one peak. However, the two experiments differ in run durations: GALLEX runs are either three weeks or four weeks (approximately) in duration, whereas GNO runs are all about four weeks in duration. When we form 3-week and 4-week subsets of the GALLEX data, we find that the relevant histograms are unimodal. The upper peak arises mainly from the 3-week runs, and the lower peak from the 4-week runs. The 4-week subset of the GALLEX dataset is found to be similar to the GNO dataset. A recent re-analysis of GALLEX data leads to a unimodal histogram.

## 1. Introduction

It has been known for some time that the published capture-rate measurements derived from the GALLEX solar neutrino experiment are bimodal in the sense that the histogram formed from these data has two distinct peaks (Sturrock and Scargle, 2001). We recently introduced a "bimodality index" to provide an objective measurement of bimodality (Sturrock, 2008). This analysis confirmed that the published capture rate measurements for the GALLEX experiment are indeed bimodal, but those for the GNO experiment are not. It appeared that the explanation must rest either in some systematic effect that is different in the two experiments, or in time-dependence of the solar neutrino flux. The purpose of this article is to try to resolve this question.

We have recently noticed a difference between the two datasets: the GALLEX data were derived from a mixture of run durations - some about three weeks in length, and the rest about four weeks in length. By contrast, the GNO dataset is composed entirely of runs of about four weeks in length. In this article, we examine the possibility that the difference concerning bimodality may be related to the difference in run durations.

In Section 2, we show the histograms for GALLEX and GNO capture-rate measurements, and we also show histograms of the run durations for the two experiments. We also note the values of the bimodality index for the two experiments. In Section 3, we examine the capture-rate histograms and the values of the bimodality index for two subsets of the GALLEX data: those with run durations of about 3 weeks, and those with run durations of about 4 weeks or more. We find that each subset is unimodal. The upper peak in the GALLEX histogram comes predominantly from short runs, and the lower peak comes predominantly from long runs. We discuss these results in Section 4.

## 2. Histograms

We have available 65 published measurements in the GALLEX series (GALLEX I: Anselmann *et al.*, 1993; GALLEX II: Anselmann *et al.*, 1995; GALLEX III: Hampel *et al.*, 1996; GALLEX IV: Hampel *et al.*, 1999) and 58 in the GNO series (Altmann *et al.*, 2000; Kirsten *et al.*, 2003; Altmann *et al.*, 2005). Histograms formed from GALLEX and GNO capture rate values are shown in Figures 1 and 2, respectively. It certainly appears that the GALLEX histogram is bimodal, but there is no obvious evidence of bimodality in the GNO histogram.

In order to assign a significance level to the bimodality of a histogram, we have recently introduced a "bimodality index" $B$ (Sturrock, 2008). This index has the property that, if the neutrino flux is constant, we expect that $B$ will be distributed exponentially so that the probability of obtaining the value $B$ or more is $e^{-B}$. We found that $B = 0.54$ for GNO data, confirming that the dataset is unimodal. For GALLEX, $B = 8.50$: according to the conventional interpretation, we find that the probability of obtaining this value or more by chance is 0.0002. However, a "P-value," such as a power measurement, should not be interpreted as the probability that the null hypothesis is correct. This more interesting probability may be estimated by a Bayesian procedure (Sturrock and Scargle, 2009). According to this procedure, the odds that the histogram is unimodal is given by

$$\Omega = 2.5(1.65 + B)e^{-B} ; \qquad (1)$$

the probability is then given by

$$P = \frac{\Omega}{1+\Omega}. \qquad (2)$$

The probability that the count-rate histogram is unimodal is found to be 0.76 for GNO. and 0.005 for GALLEX.

## 3. Subsets

The histograms formed from the run durations are shown in Figures 3 and 4 for GALLEX and GNO, respectively. We see that these differ significantly. For GNO, most runs have durations close to 4 weeks, but for GALLEX about 45% of the runs have durations close to 3 weeks, and the remainder – 55% – have durations close to 4 weeks.

In view of the difference in the duration histograms, we divide the GALLEX runs into two groups: the "3-week" group has 29 members, one of 1.82 weeks, one of 3.55 weeks, and the rest between 2.71 and 3.13 weeks; the "4-week" group has 36 members, one of 5.84 weeks, and the rest between 3.70 and 4.17 weeks. For GNO, 51 runs have durations between 3.70 and 4.12 weeks, and 7 have durations longer than 4.12 weeks.

We show histograms of the count rate for the 3-week subset and for the 4-week subset in Figures 5 and 6, respectively. We see that both histograms appear to be unimodal. We may verify this by computing the bimodality index for each subset. We obtain the values $B = 1.97$ for the 3-week subset, and $B = 3.66$ for the 4-week subset. According to the conventional interpretation, we find that the probabilities of obtaining these values or more by chance are 0.14 and 0.026 for the 3-week and 4-week subsets, respectively. However, Equations (1) and (2) gives probabilities of 0.56 and 0.25 that the histograms are unimodal for the 3-week and 4-week subsets, respectively, indicating that there is no evidence of bimodality for either subset.

## 4. Discussion

It appears from these results that the bimodality of the histogram of count-rate measurements derived from the GALLEX experiment is associated with the fact that the duration histogram is bimodal. It therefore seems probable that the apparent bimodality is due to a systematic effect rather than to variability of the neutrino flux. If this is the case, one should probably take this fact into account in analyses of GALLEX data, such as power-spectrum analyses.

We have recently learned that GALLEX data have in fact been re-analyzed by Kaether (2007), using two procedures: one (RT) based on the "rise time" and the other (PF) on the "pulse form" of the proportional-counter signals. The corresponding histograms are shown in Figures 7 and 8, respectively. We see that the "rise-time" histogram still appears to be bimodal, but the "pulse-form" histogram appears to be unimodal. These new results appear to confirm the conclusion that the bimodality evident in the earlier GALLEX datasets was due to systematic effects in the experimental procedures rather than variability of the neutrino flux.

## Acknowledgements

Thanks are due to Jeffrey Scargle and Steven Yellin for helpful discussions of this work, which was supported by NSF Grant AST-0607572.

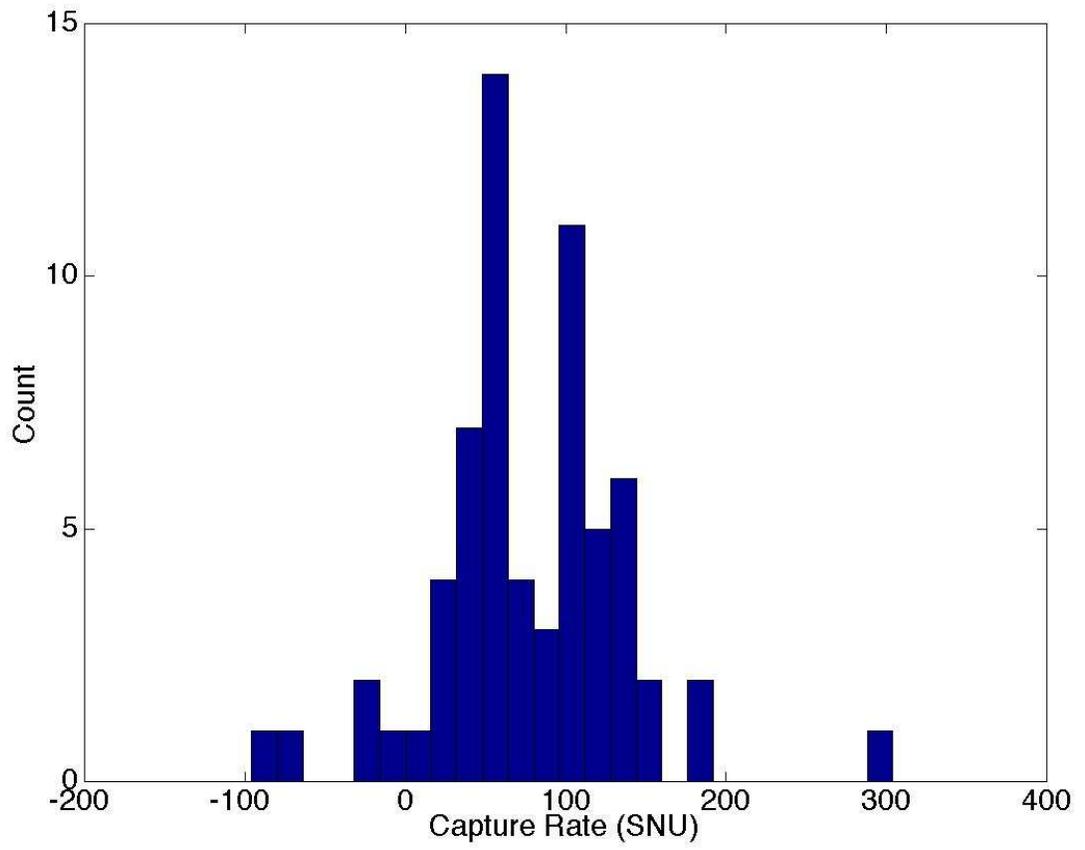

Figure 1. Histogram of GALLEX run count rate in SNU.

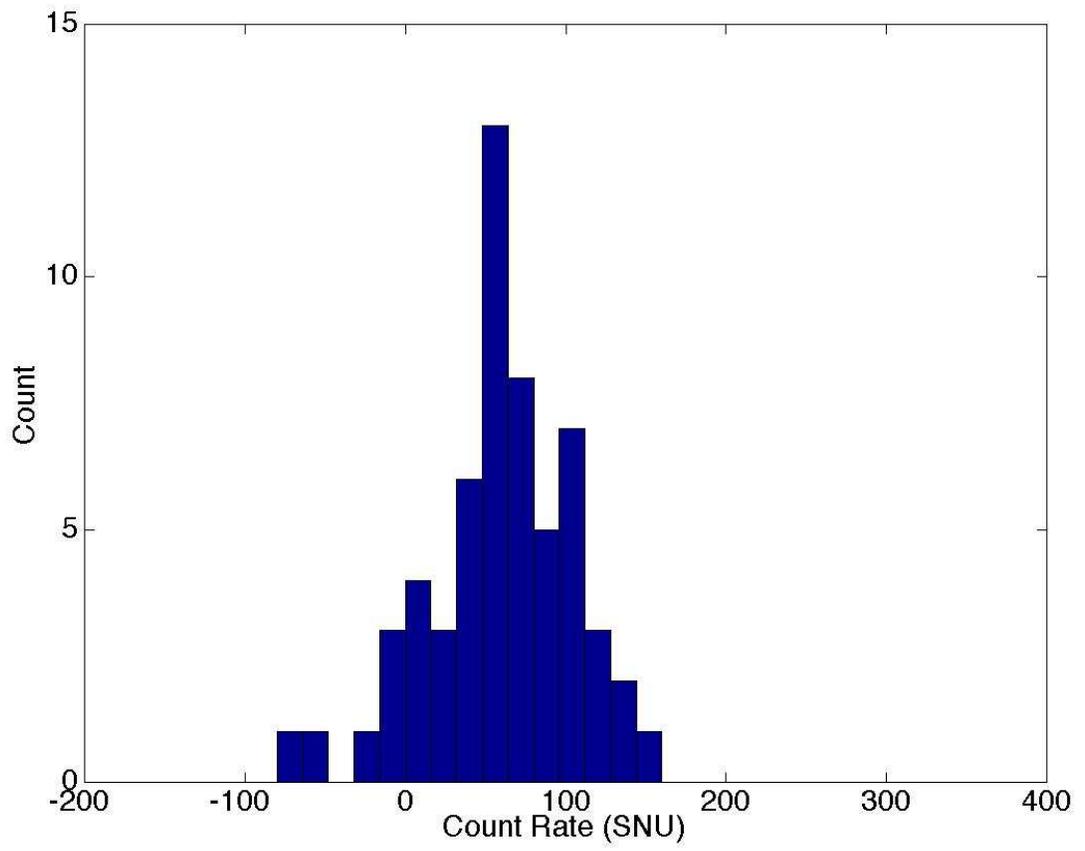

Figure 2. Histogram of GNO count rate in SNU.

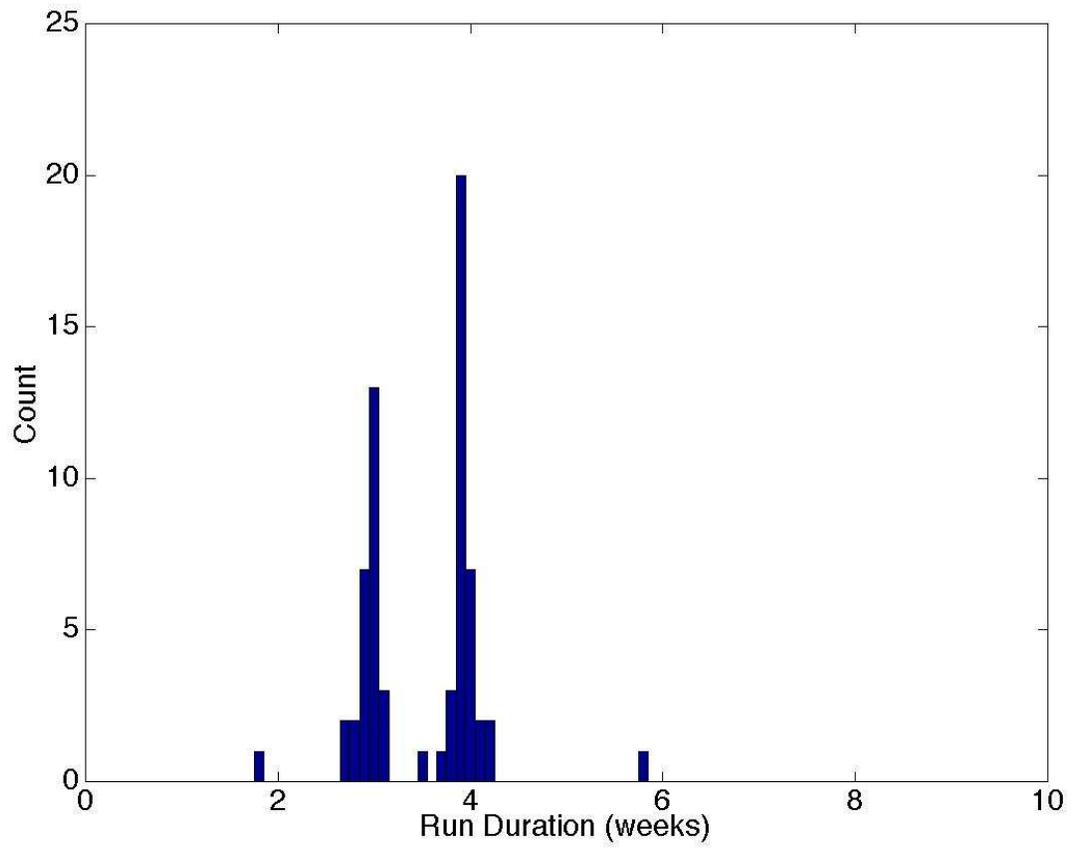

Figure 3. Histogram of GALLEX run durations in weeks.

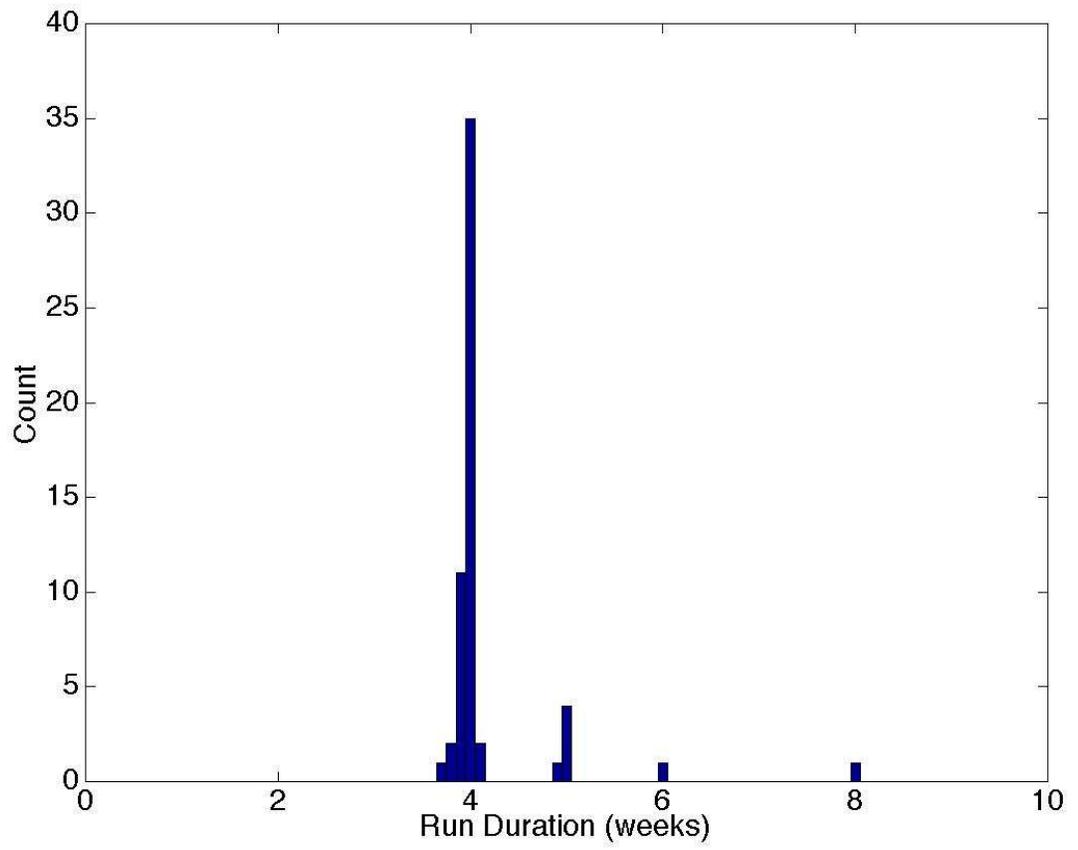

Figure 4. Histogram of GNO run durations in weeks.

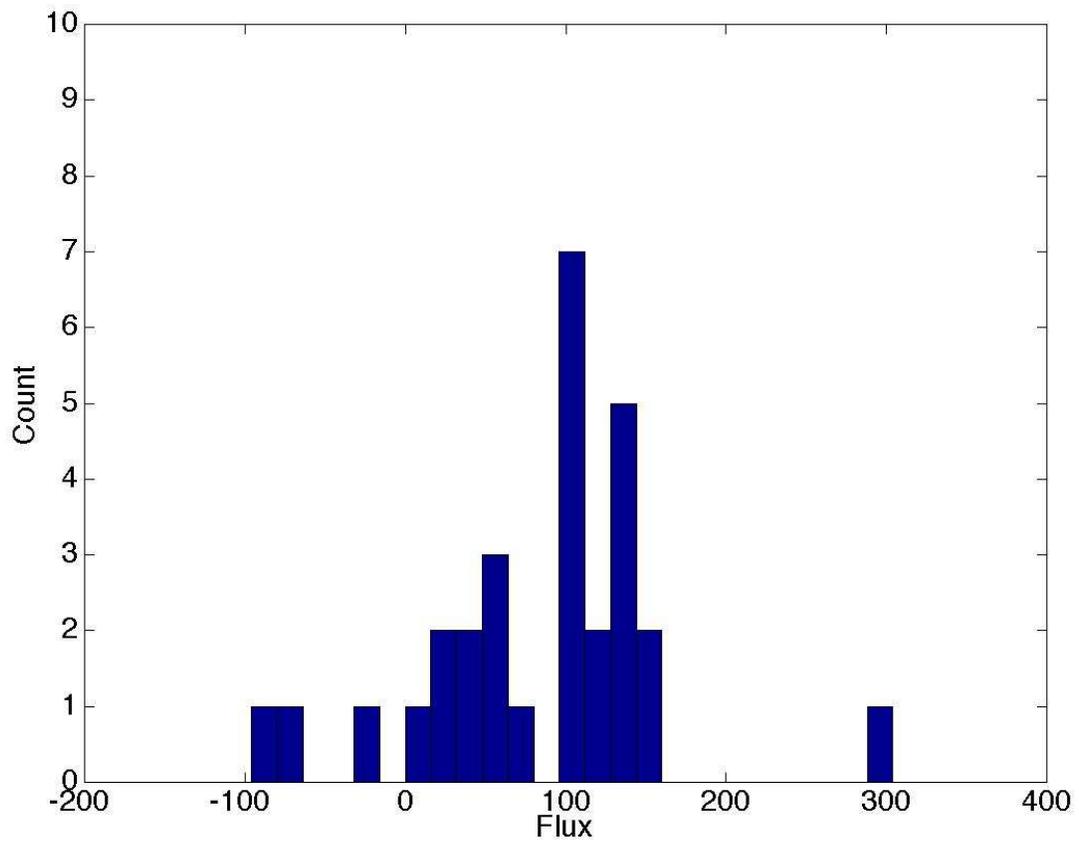

Figure 5. Histogram of count rate in SNU for the GALLEX 3-week subset.

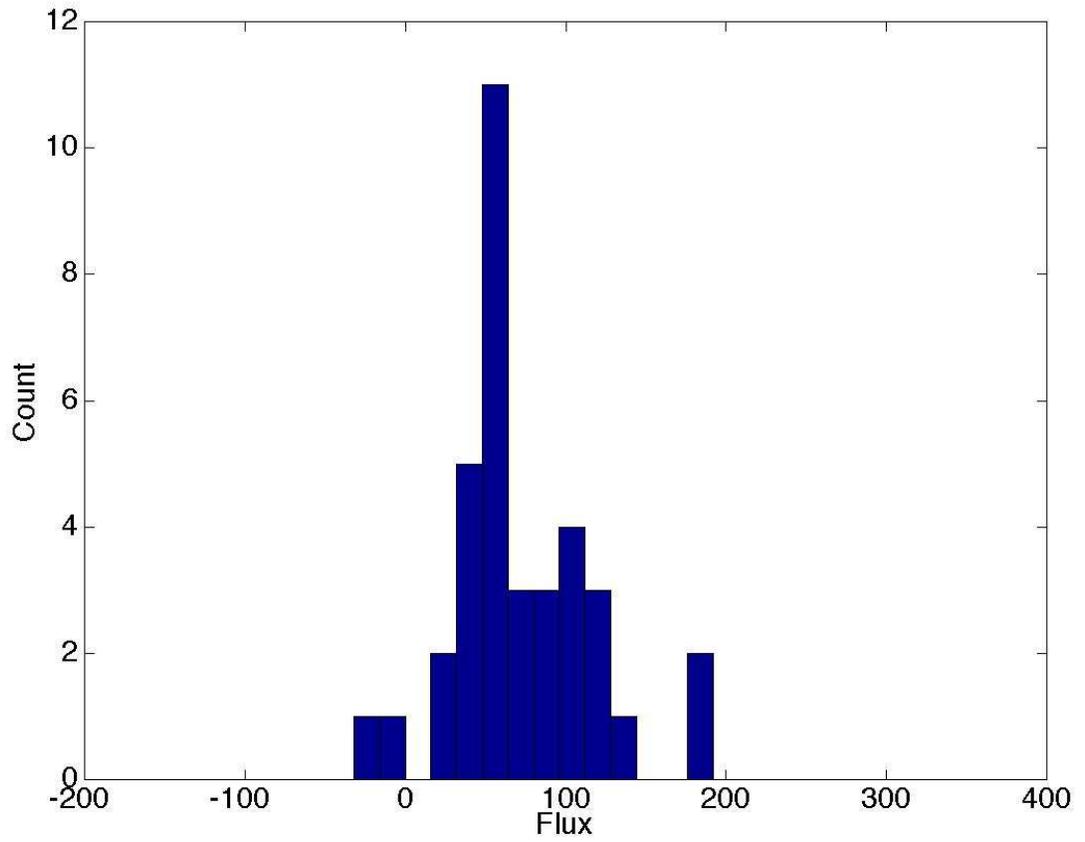

Figure 6. Histogram of count rate in SNU for the GALLEX 4-week subset.

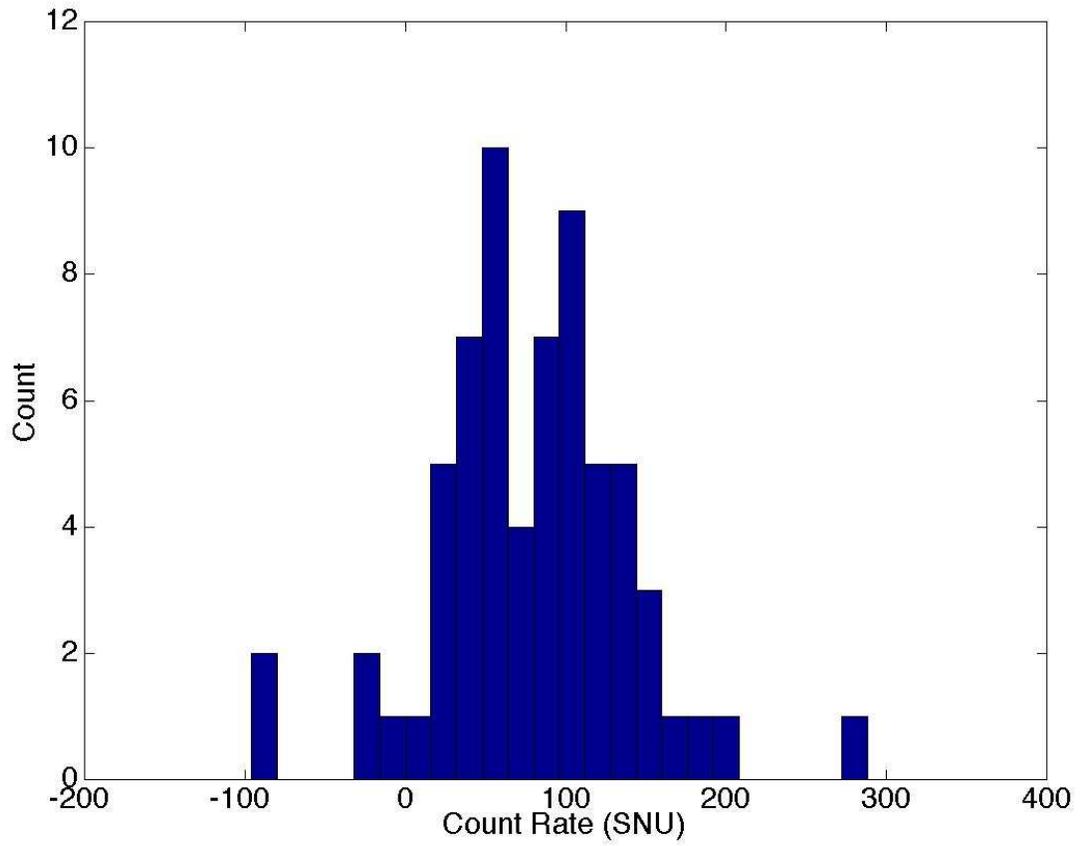

Figure 7. Histogram of the GALLEX count rate in SNU as derived from the new "rise-time" analysis.

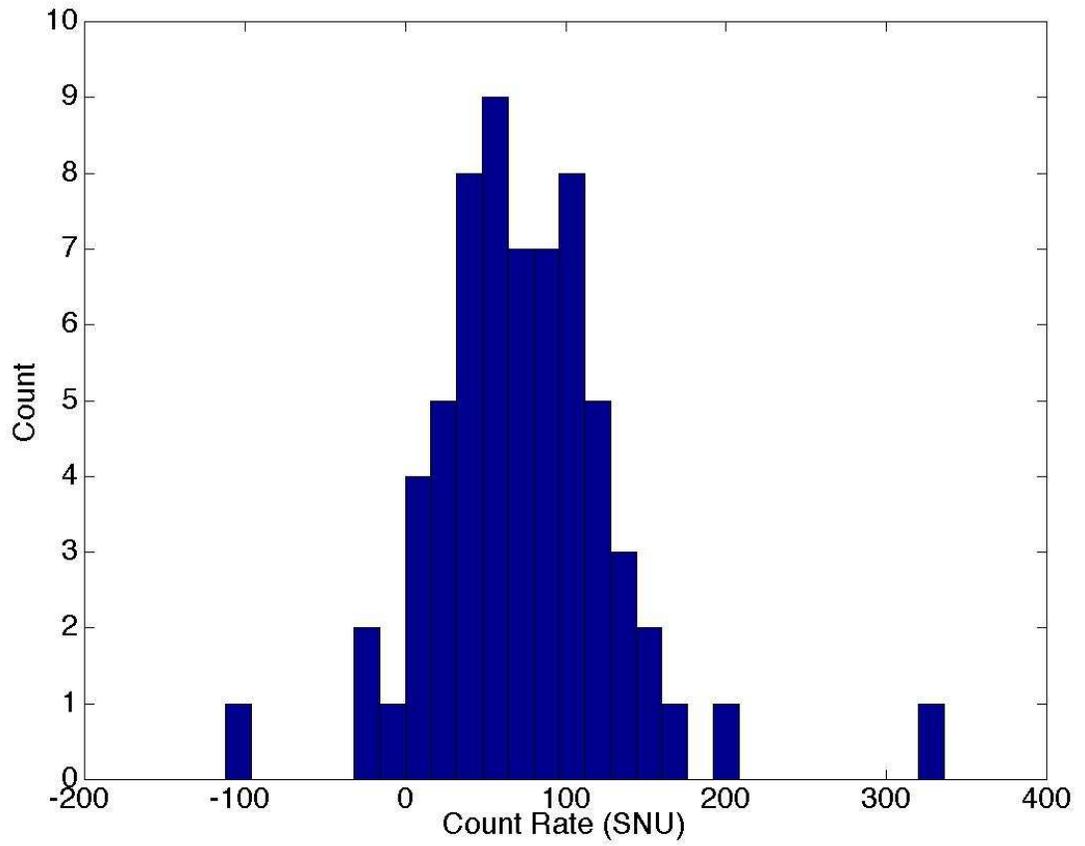

Figure 8. Histogram of the GALLEX count rate in SNU as derived from the new "pulse-form" analysis.